\documentclass[%
 reprint,
 amsmath,amssymb,
 aps,
]{revtex4-2}

\usepackage{graphicx}
\usepackage{dcolumn}
\usepackage{bm}
\usepackage{hyperref}

\usepackage{siunitx}
\usepackage{placeins}
\usepackage[normalem]{ulem}

\newcommand{\Isat}{I_{\mathrm{sat}}}
\DeclareSIUnit{\Isat}{I_{\mathrm{sat}}}

\newcommand{\G}{\Gamma}

\begin{document}
\newcommand{\orcid}[1]{\href{https://orcid.org/#1}{\includegraphics[width=10pt]{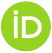}}}
\preprint{APS/123-QED}

\title{Characterisation of a strontium cold atom source using fluorescence spectroscopy and time-of-flight}

\author{Kamran Hussain$^{1,2}$\orcid{0009-0000-8148-1499}, Hamza Labiad$^{1,3}$\orcid{0009-0002-9806-2842}, Anna L. Marchant$^1$\orcid{0000-0002-6350-4842}, Jonathan N. Tinsley$^{2}$\orcid{0000-0002-4925-9350}, Tristan Valenzuela$^1$\orcid{0000-0002-5402-6615}, Jonathon Coleman$^2$\orcid{0000-0003-1319-0889}, Dave Newbold$^{4}$ and Mark G. Bason$^{1}$\orcid{0000-0003-1921-524X}}
\email{mark.bason@stfc.ac.uk}
\affiliation{%
 $^1$  RAL Space, Rutherford Appleton Laboratory, STFC, Harwell Campus,
Didcot, OX11 0QX, UK}%

\affiliation{%
 $^2$ Department of Physics, University of Liverpool, Liverpool, Merseyside, L69 7ZE, UK}%

\affiliation{%
$^3$ National Quantum Computing Centre, Rutherford Appleton Laboratory, STFC, Harwell Campus,
Didcot, OX11 0QX, UK}%

\affiliation{
$^4$Rutherford Appleton Laboratory, UKRI-STFC, Harwell Campus,
Didcot, OX11 0QX, UK}

\date{\today}

\begin{abstract}
We demonstrate a characterisation methodology for a strontium atomic beam, produced by a two-dimensional magneto-optical trap and delivered via a resonant push beam, using fluorescence spectroscopy and time-of-flight (ToF). This provides insight into the beam characteristics of a cold atom source, allowing for direct measurement of the transverse velocity spread, longitudinal velocity distributions, divergence, and the capturable flux for further cooling. From the ToF measurements, we derive a series of flux-per-longitudinal-velocity distributions at varying push saturation parameters ($s_{\mathrm{push}}$) using both a unidirectional and counter-propagating resonant probe beam. A simulation-derived factor is applied to the unidirectional probe longitudinal velocity distribution to account for differences in the scattering rate scaling. The distributions are integrated up to an estimated 3D-MOT capture velocity of \SI{30}{\meter\per\second}. For our system, we find that at $s_{\mathrm{push}} = 0.45$, we obtain a flux of $(1.7 \pm 0.4)\times10^{8}$~atoms/s and $(1.5 \pm 0.4)\times10^{8}$~atoms/s, using a unidirectional probe beam and counter-propagating probe, respectively. These measurements provide a framework for characterising cold atomic sources for applications such as 3D MOT loading and atom interferometers.
\end{abstract}

\maketitle


\section{\label{sec:intro}Introduction}
Atom interferometers are a rapidly advancing class of quantum sensors to explore fundamental physics, such as gravitational wave detection \cite{abe_matter-wave_2021, ELGAR, ZAIGA}, dark matter searches \cite{zhou_ytterbium_2024, du_atom_2022}, inertial sensors for equivalence principle tests \cite{asenbaum_atom-interferometric_2020, aguilera_ste-questtest_2014, Hannover_EP}, and determination of fundamental constants such as the fine structure \cite{parker_measurement_2018, morel_determination_2020}. The applicability of these quantum sensors extends to navigation \cite{Geiger2020HighaccuracyIM}, gravity-gradient mapping \cite{gravity_cartography} and magnetometry \cite{spacemagnetometrydifferentialatom}.
\\~\\
Alkaline-earth atoms, such as strontium, are particularly advantageous for atom interferometry due to their narrow optical clock transitions, enabling single-photon interferometry schemes \cite{hu_sr_2020, AION_gradiometer}. The Atom Interferometry Observatory and Network (AION) \cite{AION_paper} and the Matter-wave Atomic Gradiometer Interferometric Sensor (MAGIS) \cite{abe_matter-wave_2021} are projects that aim to employ strontium-atom interferometry for gravitational wave detection in the mid-band frequency regime (0.1 - \SI{10}{\hertz}) \cite{MAGIS_gravwave_theory} and for probing the nature of dark matter, particularly ultralight dark matter (ULDM) searches in the mass range between 10$^{-17}$ - 10$^{-12}$ eV \cite{AI_DM_detector}. These initiatives involve the building of scalable atom interferometers, progressing from table-top experiments to long-baseline terrestrial sensors, and ultimately towards space-based platforms \cite{AEDGE}. The broader scope includes a network of atom interferometry experiments \cite{abend_terrestrial_2024}, represented by AION \cite{AION_paper}, MAGIS \cite{abe_matter-wave_2021}, ELGAR \cite{ELGAR}, and ZAIGA \cite{ZAIGA}. In conjunction with gravitational-wave observatories such as LIGO and VIRGO \cite{PhysRevD.101.124013}, long-baseline atom interferometers promise substantial opportunities for multi-messenger astronomy.
\\~\\
Enhancing the sensitivity of atom interferometers to ULDM and for mid-band gravitational wave detection requires further technological advancements. These include the development of high-power laser sources \cite{derose_high-power_2023}, large momentum transfer \cite{Rudolph_LMT}, longer baselines \cite{Hogan2011An}, delta-kick cooling techniques to achieve colder atomic ensembles \cite{ammann_delta_1997}, and spin-squeezing methods to reduce quantum noise \cite{salvi_squeezing_2018,li_spin-squeezing-enhanced_2023, fuderer_hybrid_2023}. Importantly, increasing the atomic flux is critical for both maximising the atom number per cloud ($N_a$) and reducing the loading time ($\Delta t$) of the cold atom ensemble between experimental runs. The higher atom count and repetition rate would increase the signal-to-noise ratio \cite{inertial_sensing_Qgases}, reducing the statistical uncertainty and lowering the smallest resolvable ULDM coupling, which scales as $d^{\textrm{best}}_{\phi} \sim \sqrt{\frac{\Delta t}{N_{a}}}$~\cite{badurina_refined_2022}.
\\~\\
In the pursuit of high-flux atomic sources, reliable and accurate characterisation techniques are essential for continuously improving source design and comparing atomic sources with each other. In this work, a cooled strontium atomic beam is generated by a two-dimensional magneto-optical trap (2D MOT) and delivered into a measurement chamber using a resonant push beam. First, we characterise the effect of a unidirectional fluorescence probe beam on atoms and show that its scattering force shifts the atomic cloud, yielding asymmetric spectra that are blue-shifted and broadened, in agreement with simulated spectra. To mitigate these probe-induced distortions in the atomic cloud, we compare it with a counter-propagating beam by retro-reflecting the probe and observe a reduction in both asymmetry and spectral broadening. We then extract the atomic flux distribution over a range of longitudinal velocities at different push beam intensities using a ToF method, with both probing schemes, scaling the unidirectional-probe flux distributions by a simulation-derived factor. Using a combination of ToF and fluorescence spectroscopy, the most probable transverse and longitudinal velocities can be measured, providing insights into the atomic beam dynamics, divergence, and cloud diameter at the centre of the chamber as a function of the push beam saturation parameter. These measurements provide a means of optimising the flux for 3D MOT loading and quantifying the atomic cloud ensembles for atom interferometry.

\section{\label{sec:exp_app}Experimental apparatus}
\begin{figure}[!ht]
            \includegraphics[width=\columnwidth]{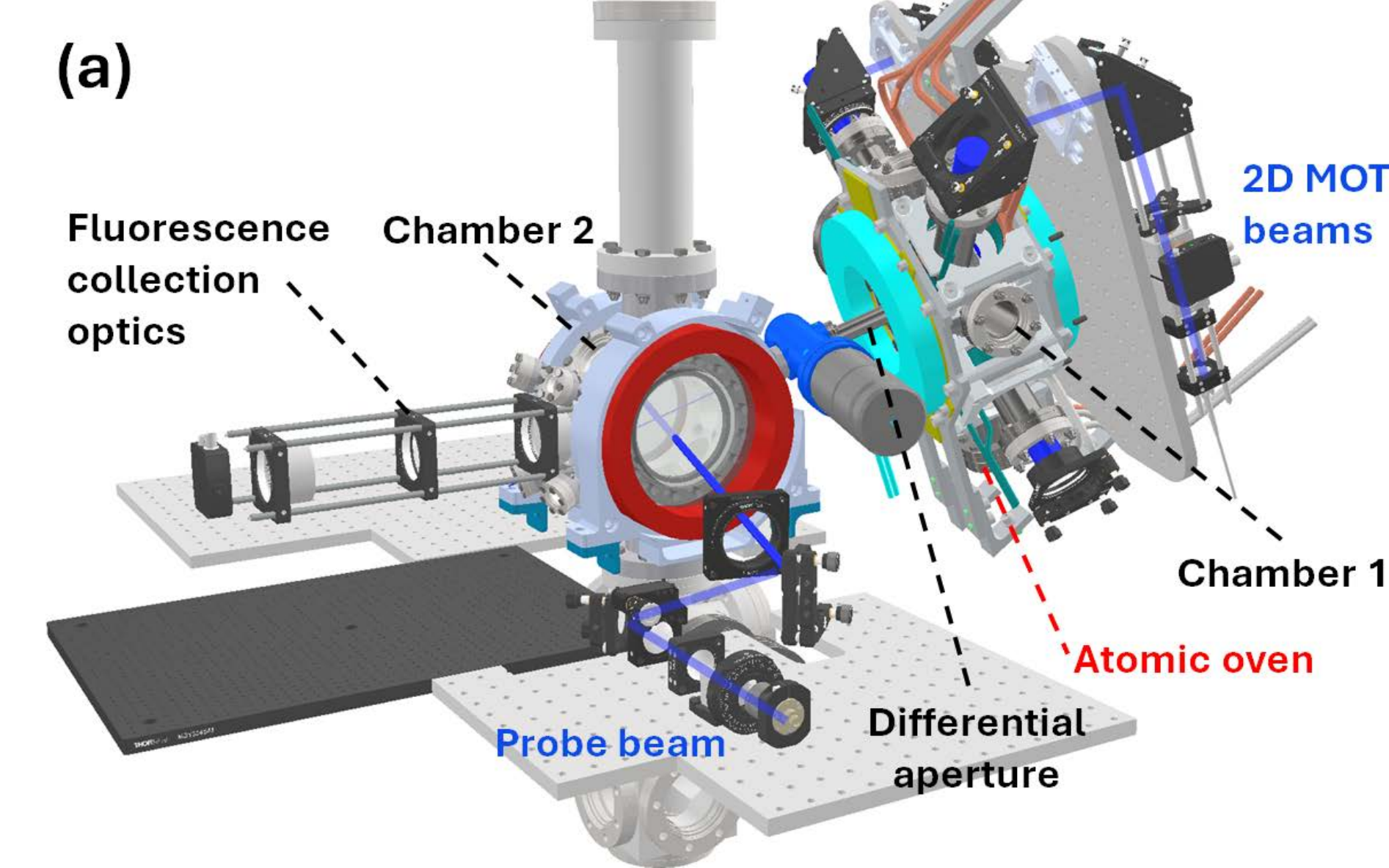}
            \includegraphics[width=\columnwidth]{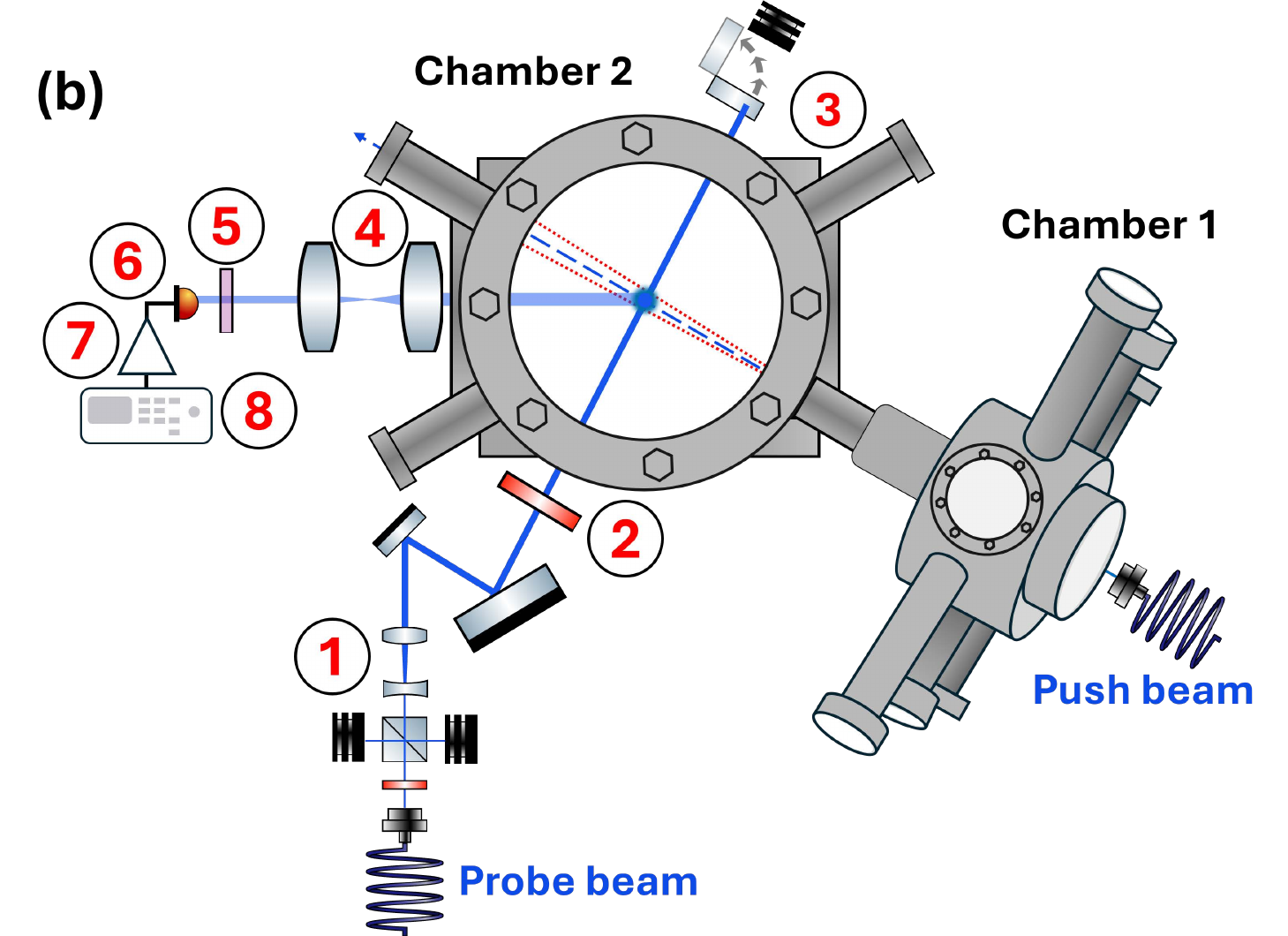}
	\caption{(a) CAD model of the two-chamber vacuum system. Chamber 1 provides the source of cold strontium atoms for the experiment via an atomic oven, using a 2D MOT configuration. The atoms are then transferred into chamber 2 by a push beam, where they are characterised using fluorescence spectroscopy and time-of-flight. (b) The top-down layout of the experimental setup for characterising the atomic beam via a transverse probe beam. The layout contains the following: (1) Galilean beam expander pair, (2) Half-wave plate, (3) Retro-reflection mirror, (4) Fluorescence imaging lens-pair, (5) Bandpass filter centred at \SI{461}{\nano\meter}, (6) Si-photodiode, (7) Transimpedance amplifier, (8) Oscilloscope.}
	\label{fig:Experimental chamber and setup}
\end{figure}

The apparatus, shown in figure~\ref{fig:Experimental chamber and setup}(a), and described in further detail in Ref.~\cite{AION_sidearm}, makes use of two ultra-high vacuum chambers, separated by a cylindrical differential aperture of length \SI{15}{\milli\meter} and diameter \SI{3}{\milli\meter}. An atomic oven heated to around \SI{400}{\celsius} generates a flux of hot strontium atoms upwards into chamber 1, where atoms are cooled transversely using a 2D MOT based on laser-cooling on the broad ($\G = 2\pi\times$\SI{30.29} {\mega\hertz}) $^1$S$_0$$-$$^1$P$_1$ \SI{461}{\nano\meter} transition of $^{88}\textrm{Sr}$. Permanent neodymium magnets along the chamber arms generate a magnetic-field gradient of \SI{37}{G\per\centi\meter}, producing a field zero at the chamber centre for the formation of a 2D MOT, with bias coils providing fine alignment to the differential-aperture centre. Atoms are delivered into chamber 2 via a resonant push beam, where they will ultimately be loaded into a 3D MOT. In this work, we instead use chamber 2 to characterise the atomic beam using fluorescence spectroscopy and time-of-flight across different push and probe beam parameters, with the probe and imaging optics illustrated in figure~\ref{fig:Experimental chamber and setup}(b).
\\~\\
Two laser systems are used to conduct the experiment. The 2D MOT and push beams are provided by high-power diode lasers (Nichia NDB4916E), which are injection-locked to a \SI{461}{\nano\meter} Extended-Cavity Diode Laser (ECDL) that is frequency-stabilised to a hollow-cathode lamp via a polarisation-spectroscopy locking scheme~\cite{Shimada_2013}. The probe beam source is provided by a separate free-running laser system referenced to the ECDL via a beat note, which is read out with a frequency counter (Keysight 53220A). The probe beam frequency is controlled by varying the piezo voltage applied to the laser's internal cavity. The voltage is adjusted such that the probe beam covers a frequency range of \SIrange{-3}{3}{\G} from the observed resonance frequency in chamber 2. The frequencies for each laser system are monitored on a wavemeter (HighFinesse WS8) calibrated to a HeNe laser.
\\~\\
The atomic beam is imaged at the centre of chamber 2, where a probe beam is aligned perpendicular to the atomic beam propagation, addressing the transverse velocity. Fluorescence is collected using a pair of lenses $f_1 = $~\SI{150}{\milli\meter} and $f_2 = $~\SI{75}{\milli\meter} on a Si-photodetector with a bandpass filter centred at \SI{461}{\nano\meter}, and the signal is amplified using an adjustable high-gain transimpedance amplifier (FEMTO DLPCA-200). The direction of the light scattered from the cloud depends on the polarisation of the probe beam, see Appendix~\ref{Appendix:DipoleEmission}. 
\\~\\
Fluorescence spectroscopy provides information on the transverse velocity of the atomic beam, which is influenced by the 2D MOT dynamics and limited by the differential aperture, whereas the longitudinal velocity is largely determined by the push beam intensity. Understanding the longitudinal velocity as a function of the push parameters is important to maximise the capture of atoms into a 3D MOT. Fluorescence on the \mbox{$^1$S$_0-^1$P$_1$} transition at \SI{461}{\nano\meter} is governed by the two-level scattering rate given a peak probe intensity, $I_0$, and probe detuning, $\Delta$. The scattering rate is expressed as:

\begin{equation}
  \G_{\mathrm{sc}}(\Delta, I_0) = \frac{\G}{2}\,\frac{s}{1 + s + 4\Delta^2/\G^2},
  \qquad s \equiv \frac{I_0}{\Isat}
  \label{eq:scatt-rate}
\end{equation}
where $s$ is the saturation intensity ratio with $\Isat$ being the saturation intensity of the transition and $\G$ is the transition linewidth.  The parameters for the probe and push beams are given in table~\ref{beam_param}.

\section{\label{sec:characterisation}Cold atom source characterisation}
\begin{figure}[!ht]
        \includegraphics[width=\columnwidth]{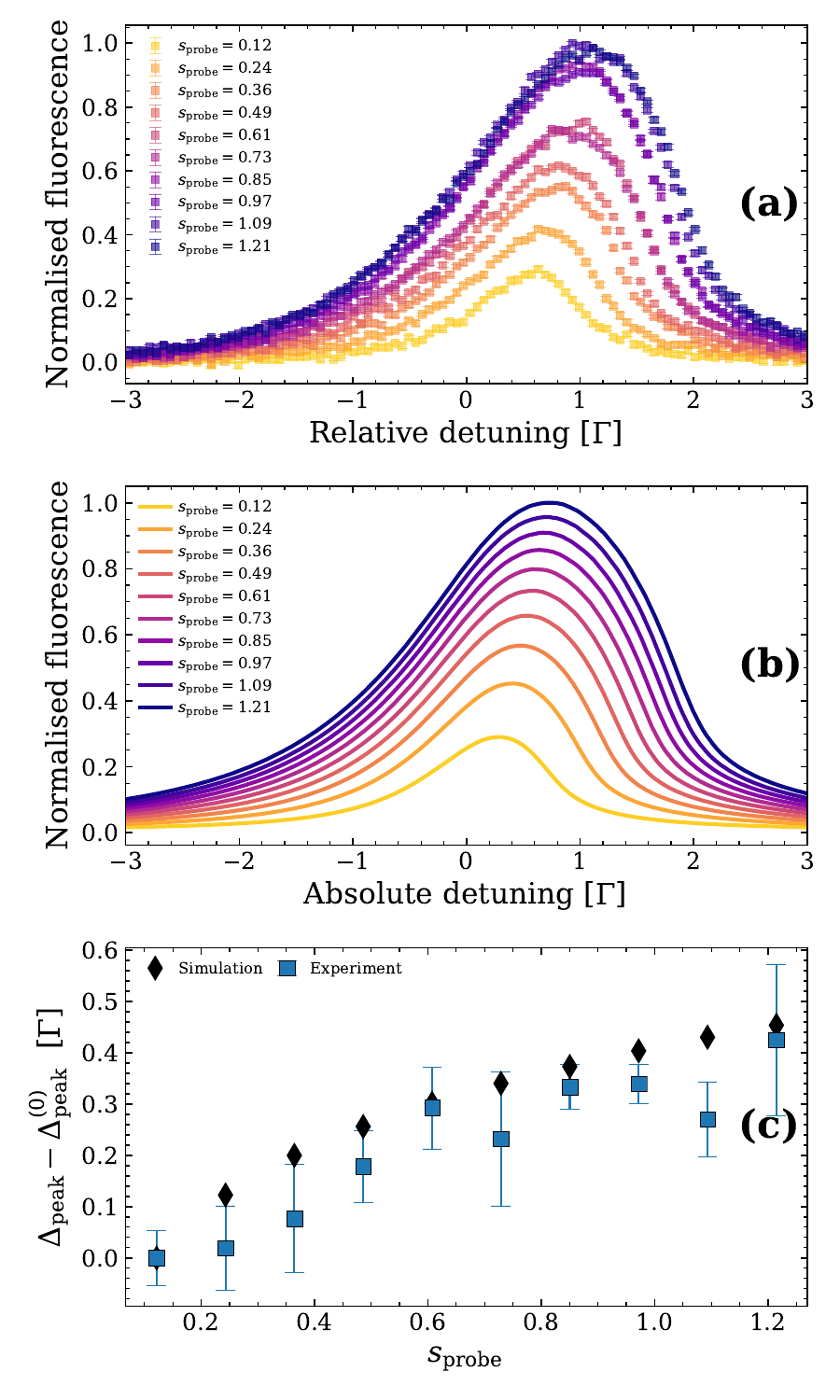}
	\caption{Transverse fluorescence spectroscopy profile of the atomic beam measured in chamber 2 using a unidirectional probe beam at peak probe saturation between $s_{\textrm{probe}}=0.12$ and $s_{\textrm{probe}}=1.21$. The probe beam frequency is scanned between \SIrange[]{-3}{3}{\textit{$\G$}} of the $^1$S$_0$$-$$^1$P$_1$ transition. (a) The fluorescence signal captured from the photodiode, of a single measurement, with the relative detuning axis calculated using equation~\ref{eq:detuning_scale} based on the difference between the beat note frequency and spectroscopy lock frequency. Data points are binned (100 evenly spaced bins across the scanned detuning range) for display clarity. (b) Simulated spectra at the same $s_{\textrm{probe}}$ as the experiment. (c) Comparison between the peak detuning ($\Delta_{\mathrm{peak}}$) relative to the respective lowest-intensity peak ($\Delta_{\mathrm{peak}}^{(0)}$) for the experimental (blue square) and simulation (black diamond) spectra to highlight the intensity-dependent shift.}
	\label{fig:Spectra profile - probe beam}
\end{figure}

\begin{figure}[!ht]
        \includegraphics[width=\columnwidth]{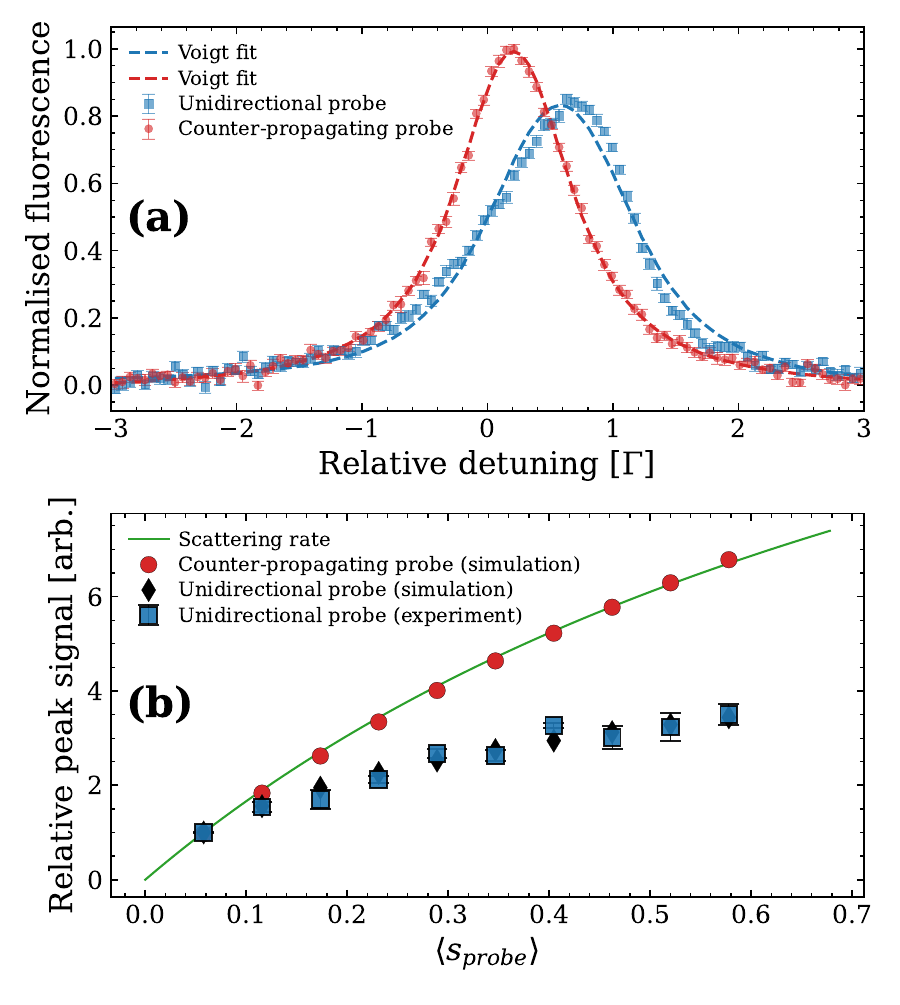}
	\caption{Transverse spectroscopy profile comparison between a unidirectional and retro-reflected counter-propagating probe configuration at the same push beam saturation ($s_{\textrm{push}} = $~0.22), taken between \SIrange[]{-3}{3}{\textit{$\G$}} of the $^1$S$_0$$-$$^1$P$_1$ transition. (a) Experimental data of the spectroscopy profile for a unidirectional probe beam (blue) and a retro probe beam (red), both at an average saturation parameter $\langle s_{\textrm{probe}}\rangle = $~0.12. Data points are binned (100 bins across the scanned detuning range) for display clarity; Voigt fits are performed on the unbinned data. (b) Relative peak fluorescence signal as a function of average probe intensity seen by the atoms, comparing the unidirectional probe measurement (blue) with the corresponding unidirectional probe simulation (black) and the counter-propagating probe simulation (red). The solid curve (green) shows the expected peak signal from the two-level scattering-rate formula. The data points are all normalised to the lowest probe intensity.} 
	\label{fig:Spectra profile - probe comparison}
\end{figure}

\subsection{Fluorescence spectroscopy}
To explore the effect of probe intensity on the measured fluorescence signal, we first align the single probe beam perpendicular to the atomic beam and vary the beam power, recording the fluorescence at the chamber centre. The resulting spectroscopy signal of a unidirectional probe beam can be seen in figure~\ref{fig:Spectra profile - probe beam}(a) as it evolves with increasing probe saturation parameter ranging from $s_{\textrm{probe}} = $~\SIrange{0.12}{1.21}{\textit{}}, showing linewidth broadening and an increasingly asymmetric spectral profile. Above the saturation intensity ($s_{\textrm{probe}} \geq 1$), we expect the linewidth to increase due to power broadening; however, power broadening alone should not shift the spectral peak centre. Instead, we attribute the observed peak shift and asymmetry to the scattering force from a unidirectional probe beam, accelerating the atomic cloud along the fluorescence beam-propagation axis, thereby imparting a net velocity to the atoms away from the probe, resulting in a blue shift.
\\~\\
The expected magnitude of this effect can be estimated using the scattering-rate model (equation~\ref{eq:scatt-rate}), and the discrepancy between the predicted and measured shifts is investigated using a Monte Carlo simulation to estimate the radiation pressure on the atomic beam, which in turn is used to estimate the scattering rate and fluorescence signal. It is based on a methodology described in detail in Ref.~\cite{bandarupally_design_2023}, which calculates the force from a unidirectional beam on a two-level atom at discrete time points, updating the position and velocity after each step. The force and number of scattered photons at each step depend on the laser beam intensities, polarisation and directions at the current atom position, which is assumed to be constant during each time step. The effect of Doppler heating is also approximated via a pseudo-random walk. The simulation accounts for experimentally measured parameters of the push and probe laser beams, as well as the measured velocity distributions. The chosen time steps (\SI{20}{\micro\second} steps) are considerably larger than the excited state lifetime, and we estimate losses via $^1$P$_1$$-$$^1$D$_2$ to be negligible for our beam powers and transit times, permitting this semi-classical treatment as a two-level $^1$S$_0-^1$P$_1$ system~\cite{Hanley30032018}.
\\~\\
The simulations were performed for 10$^4$~atoms for a range of probe beam detunings at each experimental probe beam intensity, with the total number of scattered photons estimated in each case. This includes accounting for the field of view of the collection optics and the dipole emission of the atoms. The simulations exhibit the same change in shape and shift in peak fluorescence as a function of detuning with increasing probe beam intensity, as well as very well approximating the increase in total collected photons with increasing beam intensity (figure~\ref{fig:Spectra profile - probe beam}(b)). However, there is a clear offset between the frequency axis of the simulation and the experimental data, especially at higher probe saturation. This could be the result of frequency drift, with the absence of an independent measurement of the ECDL frequency making a direct comparison of these two data sets difficult. However, both the experiment and the simulation exhibit asymmetric profiles and an associated shift toward higher (bluer) detunings. To focus on the relative intensity-dependent shifts, the simulated and experimental peak detuning shifts (from each respective first peak) for each probe saturation parameter are plotted in figure~\ref{fig:Spectra profile - probe beam}(c), and the observed trend shows good agreement within the error bars between the experiment and the simulation.
\\~\\
Introducing a retro-reflected counter-propagating  probe balances the radiation-pressure force, thereby suppressing the intensity-dependent asymmetry and detuning offset observed with a unidirectional probe beam. This improves agreement with a Voigt lineshape model, enabling the Gaussian (Doppler) component of the fit to be used to extract a more reliable transverse velocity. Figure~\ref{fig:Spectra profile - probe comparison}(a) highlights the effect of retro-reflection: compared to the unidirectional probe case, the retro-probe spectrum is more symmetric about the line centre and is well described by a Voigt profile across the measured detuning range. The results of the simulation in figure~\ref{fig:Spectra profile - probe comparison}(b) further support this interpretation, showing that the counter-propagating probe peak fluorescence follows the expected saturation dependence from the two-level scattering-rate model, whereas the unidirectional probe response increasingly deviates at higher saturation intensities due to radiation-pressure acceleration and the resulting distortion of the measured spectrum.

\begin{figure*}
    \includegraphics[width=\textwidth]{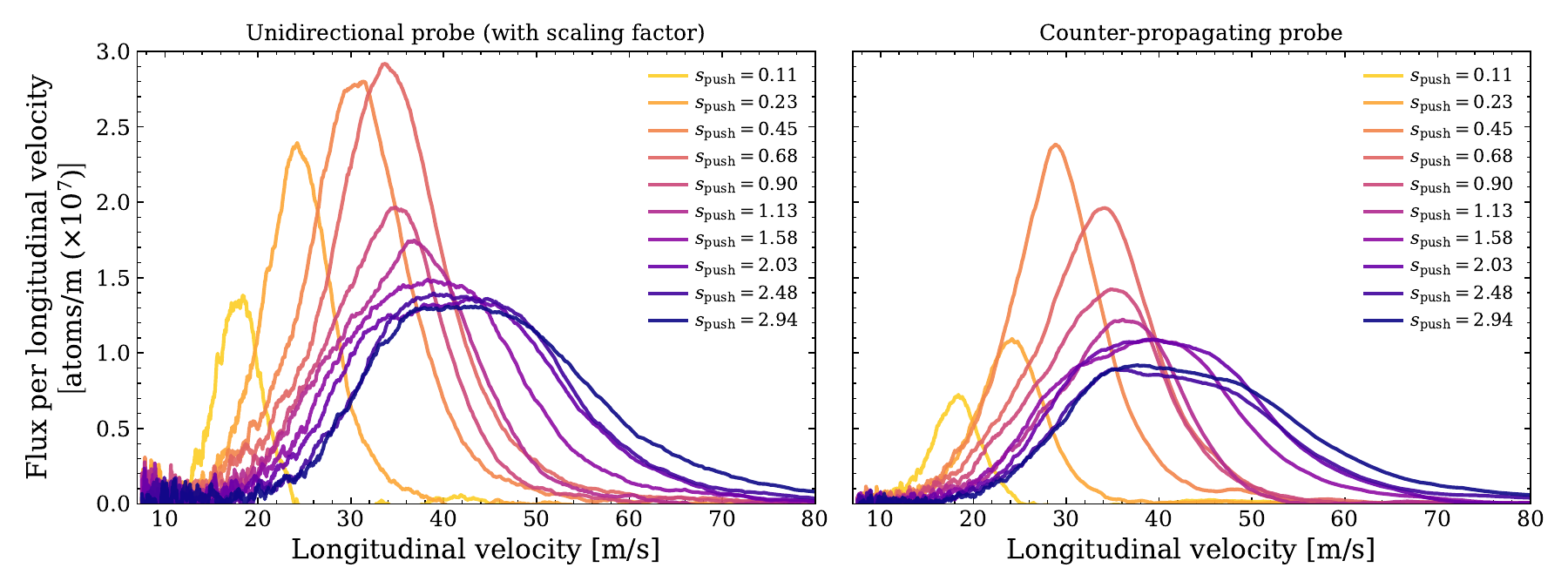}
	\caption{The fluorescence decay from the time-of-flight signal (figure~\ref{fig:TOF_decay}) is converted into units of atoms/s/m$s^{-1}$ using equation~\ref{eq:flux-per-meter} and plotted as a function of the longitudinal velocity.  The signal processing methodology is described in Appendix~\ref{Appendix:tof_processing}. The resulting distribution illustrates how the atomic beam flux per longitudinal velocity class and the most probable longitudinal velocity evolve with the push beam saturation parameter. The ToF dataset was taken using both a unidirectional probe beam (left) and a counter-propagating probe beam (right), with the scaling between the probes being influenced by a weight factor taken from the simulation shown in figure~\ref{fig:Spectra profile - probe comparison}(b). } 
	\label{fig:TOF distribution}
\end{figure*}

\begin{figure}[!ht]
        \includegraphics[width=\columnwidth]{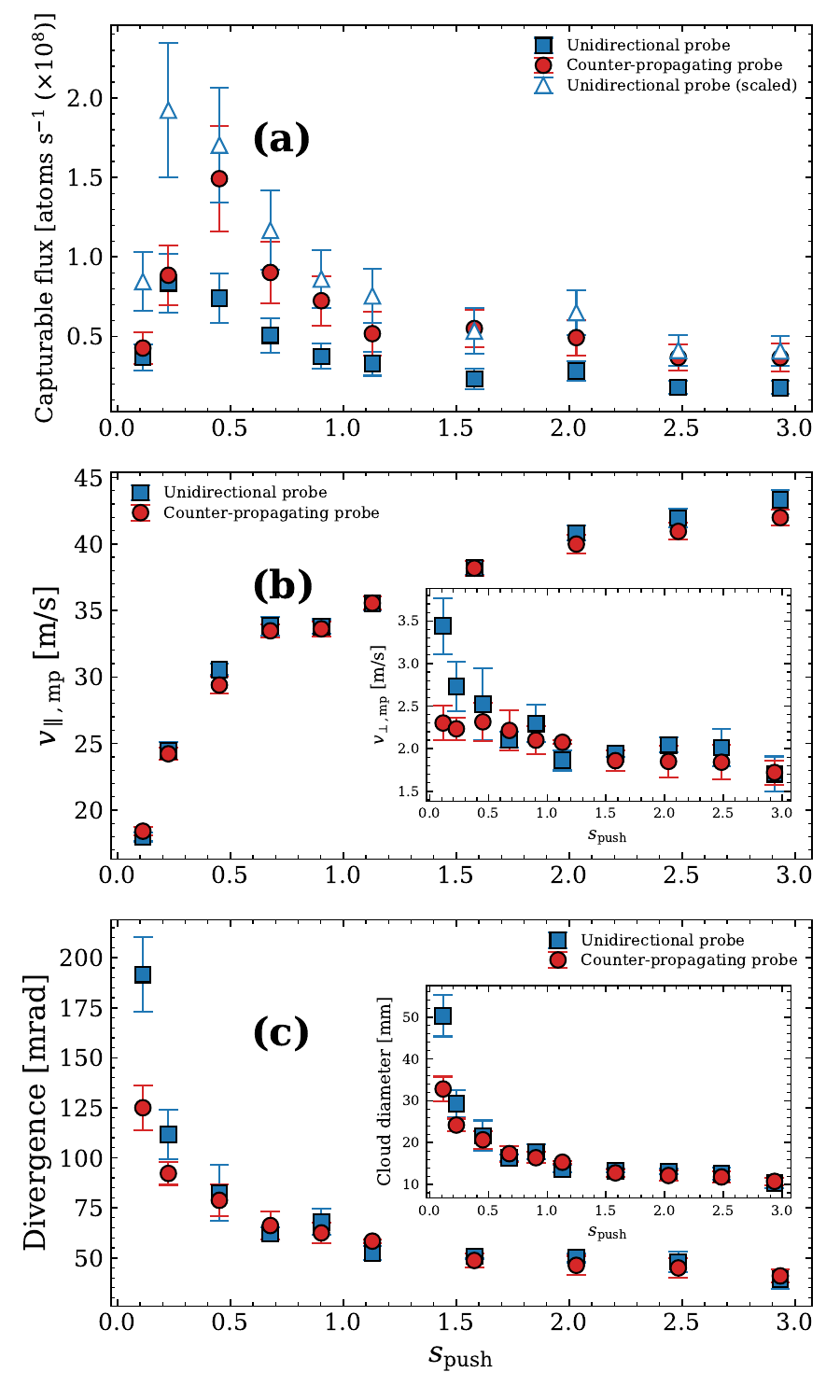}
        \caption{(a) The longitudinal atomic flux, by integrating the distributions in figure~\ref{fig:TOF distribution} for a unidirectional (filled blue square) and counter-propagating probe (red circle) between the longitudinal velocities from \SIrange[]{0}{30}{\meter\per\second} to estimate the maximum flux that can be captured by our 3D MOT and is plotted as a function of the push beam saturation to find the ideal push beam parameter. The unidirectional probe flux distribution was scaled using the simulation results in figure~\ref{fig:Spectra profile - probe comparison}(b). The unidirectional probe flux without scaling (unfilled blue triangle) illustrates this importance. (b) The evolution of the most probable longitudinal velocities is evaluated by fitting the distributions to a Gaussian model to extract the peak velocity. The transverse velocity (inset), taken from independent fluorescence spectroscopy measurements, using the same push saturation range as the time-of-flight but with $s_{\textrm{probe}}$ = 0.11. The spectra are then fitted to a Voigt model, and the velocity is extracted from the Doppler width (Gaussian). (c) The atomic beam divergence, calculated by taking the ratio of $\frac{v_{\perp}}{v_{\parallel}}$, and the atomic cloud diameter (inset) at different push saturations.}
	\label{fig:Atomic_beam_characteristic}
\end{figure}

\subsection{Time-of-Flight (ToF)}
To obtain a direct measurement of the longitudinal atomic beam velocity, a time-of-flight (ToF) method~\cite{PhysRevLett.77.3331,PhysRevA.58.3891,PhysRevA.66.023410,tiecke_high-flux_2009,Nosske_paper} is employed, which monitors the decay of the fluorescence signal by switching off the 2D-MOT cooling beams using acousto-optical modulators (AOMs) in the beam paths, while keeping the push beam on. After the 2D-MOT beams are extinguished, atoms already in flight continue to reach the probe region and fluoresce, producing a decaying signal. Defining a transit time ($t$) for atoms to travel a distance ($\ell$) from the differential-aperture entrance to the probe beam area at the centre of chamber 2, the longitudinal velocity is expressed as $v_{\parallel} = \ell/t$, assuming a constant velocity of the atom. The ToF data were taken using both a unidirectional and counter-propagating probe, but since the probe beam is not actively frequency-stabilised, the probe frequency is adjusted to maximise fluorescence at the start of each measurement to compensate, and over the course of the \SI{35}{\milli\second} acquisition window, the frequency drift is negligible. The decay is measured for several push beam saturation parameters over $s_{\textrm{push}} = $~\SIrange{0.11}{2.94}{\textit{}}. Increasing the push beam intensity is expected to increase the photon scattering rate, but what we observe is the peak fluorescence signal increasing up until $s_{\textrm{push}} = 0.45$ and $s_{\textrm{push}} = 0.68$ for the counter-propagating and unidirectional beams, respectively, after which the signal decreases. To understand how the atomic beam distribution evolves at different push intensities, the fluorescence signal is converted into atoms per longitudinal velocity using equation \ref{eq:flux-per-meter}, resulting in the distribution seen in figure~\ref{fig:TOF distribution}. Because a unidirectional probe deviates from the two-level scattering-rate scaling assumed in the conversion factor (figure~\ref{fig:Spectra profile - probe comparison}(b)), the unidirectional ToF distributions are corrected by a simulation-derived weighting factor. This factor is the ratio of the two-level scattering rate to the simulated unidirectional-probe peak response, evaluated at the probe saturation used for the ToF measurements ($\langle s_{\textrm{probe}}\rangle = 0.72$), giving a single constant $\eta_{\mathrm{ToF}} \approx 2.3$ that is applied uniformly to all unidirectional distributions (see Appendix~\ref{Appendix:single_probe_scaling}). This places the unidirectional and counter-propagating data on a common scattering-rate basis, allowing the delivered flux from the two probing schemes to be compared directly. We can clearly see that the most probable longitudinal velocity shifts to higher values with increasing push beam saturation, with a pronounced broadening of the velocity distribution for $s_{\textrm{push}} > 1$ . We also observe that the distributions between single and retro probe configurations share a similar profile at $s_{\textrm{push}} > 1$, but differ in flux. At the lowest push saturations, the measured distribution can also be shaped by transmission through the differential stage, which removes the most divergent trajectories.

\subsection{Atomic beam characteristics}
We examine how the push beam saturation affects the atomic flux, the most probable longitudinal velocity and divergence downstream of the 2D MOT. To exploit high-flux sources of cold atoms, the atomic beam will ultimately need to be captured by a 3D MOT. This capture limit, which depends on the 3D MOT parameters, is estimated to be  $v_{c} =$~\SI{30}{\meter\per\second} for our case. The capturable atomic beam flux can be approximated by fitting the ToF distribution to a Gaussian function. The resulting fit parameter can be used to integrate the function analytically using equation~\ref{eq:gaussian_integral}.
The captured flux is then plotted as a function of the push beam saturation in figure~\ref{fig:Atomic_beam_characteristic}(a) to find the optimal push beam saturation parameter to maximise 3D MOT loading for this system, which we find to be $s_{\textrm{push}} = 0.45$ for both probe configurations. We observe that the scaled-unidirectional probe data shows the same trend and are in agreement within the error bars of the counter-propagating, but differs for the first two points, where the scaling factor overestimates the flux. This indicates that potentially more scattering events are occurring at lower $\textrm{s}_{\textrm{push}}$/longitudinal velocities, where the unscaled-unidirectional probe agrees with the counter-propagating probe.
\\~\\  
Figure~\ref{fig:Atomic_beam_characteristic}(b) shows that an increase in the push beam power leads to a general increase in the most probable longitudinal velocity $v_{\parallel,\mathrm{mp}}$ (defined in Appendix~\ref{Appendix:tof_processing}) but approaches a saturation regime at higher intensities, causing the velocity to plateau. The figure~\ref{fig:Atomic_beam_characteristic}(b) inset shows the transverse velocity extracted from the Voigt fits as a function of the push beam saturation. This measurement should be constant, as the transverse probe should be insensitive to any longitudinal velocity variation as the push beam intensity is increased; this can be seen with counter-propagating probe data, with an average velocity of \SI{2.1}{\meter\per\second}. However, the first two points of the unidirectional probe are in disagreement with the counter-propagating probe, yielding a higher transverse velocity. This is
consistent with the same underlying effect as the capturable flux in figure~\ref{fig:Atomic_beam_characteristic}(a), due to a higher number of scattering event at low $\textrm{s}_{\textrm{push}}$. The result of this on the fluorescence profile would be broader spectra and hence a higher extracted transverse velocity from the Voigt fit. 
\\~\\  
Given an understanding of the transverse and longitudinal velocity  propagation, we can express the atomic-beam divergence as $\theta_{\mathrm{div}} = v_{\perp,\mathrm{mp}}/v_{\parallel,\mathrm{mp}}$, as shown in figure~\ref{fig:Atomic_beam_characteristic}(c). The transmitted divergence is limited by the differential pumping stage, characterised by an aperture diameter $D$ and length $L$. The maximum transmitted (half-angle) divergence can be approximated as $\theta_{\mathrm{diff}} \approx D/L$; for $D=\SI{3}{\milli\meter}$ and $L=\SI{15}{\milli\meter}$, this gives $\theta_{\mathrm{diff}} \approx \SI{200}{\milli\radian}$. Consequently, at low $s_{\mathrm{push}}$ where the divergence is largest, a fraction of the slowest and most divergent atoms is expected to be clipped, reducing the measured flux in chamber~2 and potentially shifting the optimum flux towards higher push saturation.
\\~\\  
Since the transverse velocity is relatively constant, $\theta_{\mathrm{div}}$ is primarily set by the longitudinal velocity and therefore decreases as the push intensity increases. The divergence then plateaus for $s_{\mathrm{push}} \gtrsim 1.6$, indicating that the atomic beam has reached the minimum achievable divergence, although at the cost of lower deliverable flux. Using the measured divergence, we estimate the fluorescence cloud diameter using equation~\ref{eq:cloud_size}, as plotted on the inset of figure~\ref{fig:Atomic_beam_characteristic}(c). To maximise the atomic flux delivered into a 3D MOT (MOT beam diameter $\SI{18.4}{\milli\meter}$), a push saturation of $s_{\mathrm{push}} = 0.45$ is expected to be optimal. This optimum reflects the flux transmitted downstream of the differential stage, rather than the total flux emitted by the 2D MOT. At this operating point, the corresponding cloud diameter is (21 $\pm$ 1)~mm, sufficient for the experimental geometry for 3D MOT capture in this system.

\section{\label{sec:discussion}Discussion}
We have characterised a strontium 2D MOT cold-atom source by measuring the longitudinal velocity distribution, transverse velocity spread, divergence, and delivered atomic flux using fluorescence spectroscopy and time-of-flight (ToF). Integrating the ToF-derived distributions up to an estimated 3D-MOT capture velocity of \SI{30}{\meter\per\second} yields a delivered capturable flux of $\Phi_{\textrm{uni.(scaled)}} = (1.7 \pm 0.4)\times10^{8}$~atoms/s and $\Phi_{\textrm{counter-prop.}} = (1.5 \pm 0.4)\times10^{8}$~atoms/s, at $s_{\mathrm{push}} = 0.45$, providing an operating point that maximises loading into a ``blue'' 3D MOT for our apparatus, where both probe configurations agree within the error bar.
\\~\\
A key systematic identified in this work is the distortions of the fluorescence lineshape from a single probe beam, resulting in intensity-dependent asymmetry and blue shifts of the resonant peak. Employing a retro-reflected counter-propagating probe suppresses this effect and yields spectra that are well described by a Voigt model. However, if a counter-propagating beam is not used, we show that scaling the fluorescence signal from a unidirectional probe beam into flux can yield results comparable to those with a counter-propagating probe for $s_{\mathrm{push}} \geq 0.45$. 
\\~\\
The Sr 2D-MOT characterisation performed by Nosske \textit{et al.}~\cite{Nosske_paper} and Barbiero  \textit{et al.}~\cite{PhysRevApplied.13.014013} is in good agreement with the trends shown for the atomic flux as a function of the push beam intensity, with the flux maximising at a lower saturation parameter $s_{\textrm{push}} = $~0.3. However, the atomic beam divergences differ from those of Nosske \textit{et al.} and Barbiero \textit{et al.}, with a maximum divergence of $\sim$~88~mrad, resulting from a smaller differential aperture diameter and length. Additionally, the maximum atomic beam flux achieved by these systems is an order of magnitude greater. This is attributed to differences in the 2D-MOT generation stage. Firstly, the atomic ovens operate at higher temperatures (\SI{460}{\celsius}) than this system (\SI{400}{\celsius}), resulting in a higher flux of hot atoms. In addition, their systems include a Zeeman slower beam, resulting in a flux enhancement factor of 3 \cite{Nosske_thesis}. We also note that the inclusion of sidebands by Barbiero \textit{et al.} in the 2D MOT beam allows the beam to address atoms at higher velocities, increasing the maximum capture velocity limit within the 2D~MOT, resulting in an enhancement factor of 7 in the flux. Incorporating such upgrades to increase the downstream capturable flux is an important consideration for future iterations of our source design.

\section{\label{sec:outlook}Outlook}
The characterisation methods demonstrated here provide a practical route to optimise high-flux strontium 2D-MOT sources and to quantify the delivered, capturable flux under realistic operating conditions. Future considerations for this measurement will focus on reducing systematic uncertainty in the fluorescence-to-flux conversion by improving control of the probe detuning during ToF acquisition and by directly imaging the fluorescence region to measure cloud size and beam centring. These measurements would enable the probe-cloud overlap factor $g$ to be evaluated and would allow the fluorescence calibration to be benchmarked against an independent observable (e.g. absorption imaging). Finally, the same spectroscopy and ToF framework can be used to evaluate source upgrades aimed at increasing downstream, capturable flux, including additional longitudinal slowing (e.g. Zeeman slower), sidebands to address faster velocity classes in the 2D MOT and off-axis push geometries to avoid interfering with the atomic ensemble in the next cooling stage. These enhancements would improve the overall loading efficiency and sensitivity of long-baseline atom interferometers being developed for precision measurement and fundamental physics experiments.

\begin{acknowledgments}
This work has been funded and supported by the Science and Technology Research Council (STFC) Training Grant, STFC Particle Physics Department (PPD) studentship and by UKRI through the Quantum Technologies for Fundamental Physics programme from the following grants: ST/T007001/1 to the University of Liverpool and ST/W006510/1 to Rutherford Appleton Laboratory.
\end{acknowledgments}

\section*{Author declaration}
The authors listed declare no conflict of interest.

\section*{Data availability}
Data supporting the findings of this study are available from the corresponding author on reasonable request.

\newpage
\appendix

\section{Fluorescence collection and conversions}

\subsection{Dipole emission}
\begin{figure}[!ht]
		\includegraphics[width=\columnwidth]{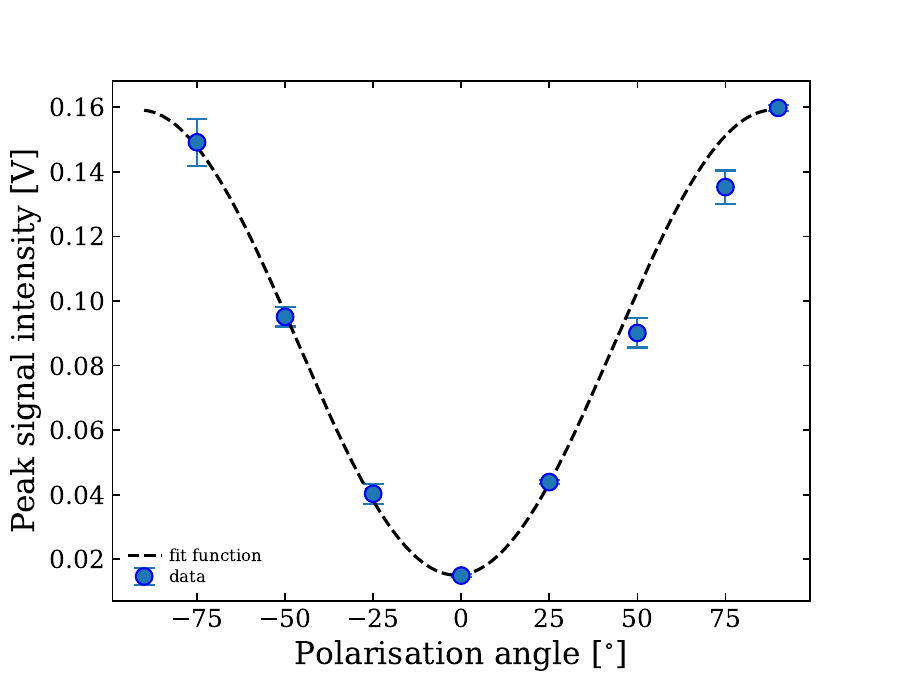}
	\caption{Measured fluorescence intensity as a function of the probe-beam linear polarisation angle adjusted using a half-wave plate along the probe axis. The data is fitted to the angular dipole emission distribution function (equation~\ref{dipole_angular_dist})}
	\label{fig:probe_polarisation}
\end{figure}  

The fluorescence from the $^1$S$_0$$-$$^1$P$_1$ transition is anisotropic and is well described by the electric-dipole radiation pattern. For a linearly driven dipole, the normalised angular distribution is:
\begin{equation} \label{dipole_angular_dist}
f(\alpha) = \frac{3}{8\pi}\mathrm{sin}^2(\alpha)
\end{equation}  
where $\alpha$ is the angle between the detection axis and the dipole axis, which is set by the probe beam polarisation \cite{Jackson1998,Griffiths2017}. The probe beam propagates perpendicular to the direction of the atomic beam (as shown in figure~\ref{fig:Experimental chamber and setup}(b)) with the detection axis $76^{\circ}$ from the probe propagation axis. Due to experimental constraints, the detection axis could not be placed at $90^{\circ}$ to the probe axis, which would maximise the detected fluorescence. Instead, a half-wave plate is placed along the probe axis to allow for the rotation of the linear polarisation axis with respect to the driven dipole axis, which modulates the detected fluorescence, producing the $\sin^2$-like trend in figure~\ref{fig:probe_polarisation}. The probes' (linear) polarisation angle is measured using a Thorlabs PAX1000VIS polarimeter.

\subsection{Solid angle and dipole emission weighting factor}

The first collection lens ($f_1 = $~\SI{150}{\milli\meter}) has a clear aperture of \SI{50.8}{\milli\meter} giving an effective radius $R=\SI{25.4}{\milli\meter}$, and is positioned a distance $D=\SI{150}{\milli\meter}$ from the chamber centre. This gives a solid angle expressed as:
\begin{equation}
\Omega_{\mathrm{PD}} =   \frac{R^2}{4D^2}.
\label{eq:Omega_PD_app}
\end{equation}  
\\~\\
We define a dimensionless dipole-emission weighting ($F_{\textrm{DE}}$) \cite{Nosske_thesis} to scale the fluorescence in accordance with the imaging system's solid angle:
\begin{equation} \label{Eq_dipole_emmision}
F_{\textrm{DE}} = \frac{3}{2\pi}\Omega^{-1}_{\textrm{PD}} \left(\arctan \left(\frac{R}{D} \right) \right)^2
\end{equation}

\label{Appendix:DipoleEmission} 

\subsection{Photovoltage-to-atom number conversion}
The atom number ($N$) is related to the photovoltage ($V_{\textrm{PD}}$) through a conversion factor ($C_{\textrm{PD}}$) \cite{Nosske_thesis} expressed as: 
\begin{equation}
N = V_{\mathrm{PD}}\,C_{\mathrm{PD}},
\qquad
  C_{\mathrm{PD}} =
\bigl(\G_{\mathrm{sc,avg}}\,\hbar\omega\,\Omega_{\mathrm{PD}}\,F_{\mathrm{DE}}\,R\,G\,T\bigr)^{-1}
  \label{eq:pd-conv}
\end{equation}
where $\G_{\textrm{sc,avg}}$ is the scattering rate defined by equation~\ref{eq:scatt-rate}, taking into account the saturation parameter, which is redefined using the average beam intensity ($I_{\textrm{avg}}$). $\hbar\omega$ is the energy of the emitted photon, $\mathcal R$ is the responsivity of the photodiode (Thorlabs DET100A2) at \SI{461}{\nano\meter}, $G$ is the gain factor of the transimpedance amplifier and $T$ is the viewport transmission. Table~\ref{tab:conversion_param} summarises the values used in the conversion factor. 
\begin{table}[h]
    \centering
    \begin{tabular}{|c|c|}
    \cline{1-2}
        Parameter & Value\\ 
        \cline{1-2}
         $\hbar \omega$ & $4.3\times10^{-19}$~J \\
         \cline{1-2}
         $\Omega_{\textrm{PD}}$ & $7.2\times10^{-3}$\\
         \cline{1-2}
         $F_{\textrm{DE}} $& $1.9$\\
         \cline{1-2}
         $\mathcal R$ & $0.25$~A/W\\
         \cline{1-2}
         $G$ & $10^{9}$~V/A\\
         \cline{1-2}
         $T$ & 92\%\\
    \cline{1-2}
    \end{tabular}
    \caption{Parameters used in the conversion factor in equation~\ref{eq:pd-conv}}
    \label{tab:conversion_param}
\end{table}

The uncertainty on the conversion factor is expressed as a percentage error,
\begin{equation}
\begin{aligned}
\frac{\Delta{C_{PD}}}{C_{PD}}
&= \biggl(
\left(\frac{\Delta_{\Gamma_{\mathrm{sc,avg}}}}{\Gamma_{\mathrm{sc,avg}}}\right)^2
+ \left(\frac{\Delta_{\Omega_{\mathrm{PD}}}}{\Omega_{\mathrm{PD}}}\right)^2 \\
&\qquad
+ \left(\frac{\Delta_{F_{\mathrm{DE}}}}{F_{\mathrm{DE}}}\right)^2
+ \left(\frac{\Delta_{\mathcal R}}{\mathcal R}\right)^2
+ \left(\frac{\Delta_G}{G}\right)^2
\biggr)^{1/2}
\end{aligned}
\end{equation}

\subsection{Flux-per-longitudinal velocity conversion}
The ToF fluorescence decay signal (figure~\ref{fig:TOF_decay}) is converted into a distribution of flux per longitudinal velocity class (figure~\ref{fig:TOF distribution}) using the following equation \cite{Nosske_paper}:
\begin{equation}
\phi(v_{\parallel}) =
\left(\frac{C_{\mathrm{PD}}}{d_{\mathrm{exc}}\cdot g}\right)\left(\frac{\ell}{v_{\parallel}}\right)\bigg(-\frac{dV_{\textrm{PD}}}{d\textrm{t}}\bigg),
\label{eq:flux-per-meter}
\end{equation}
where $d_{\textrm{exc}}$ is the diameter of the probe beam, $\ell$ is the propagation distance from the differential aperture entrance to detection region in chamber 2, $v_{\parallel}$ is the longitudinal velocity and $g$ is an expression of the spatial overlap of the probe beam with a saturation \textit{s(x,y)} over the atomic cloud distribution \textit{T(y,z)} \cite{Nosske_thesis} given as:
\begin{equation}
    g = \int_{-a}^{a} \mathrm{d}y \int_{-a}^{a} \mathrm{d}z \frac{s(x = l, y)}{s(x = l, y) + 1} \, T(y, z)
    \label{eq:gfactor}
\end{equation}
\newline
where,
\begin{equation}
    T(y,z) = \frac{1}{2\pi\sigma^2}\exp\bigg( \frac{-(y^2 + z^2)}{2\sigma^2} \bigg)
\end{equation}

The expression for $g$ is taken as a constant for the analysis ($g \approx 0.12)$.

\section{Data processing method}

\subsection{Spectroscopy processing pipeline}\label{Appendix:spectroscopy_processing}

The raw fluorescence trace, probe beam piezo input voltage, and a TTL trigger are recorded simultaneously on an oscilloscope, with the beat-note frequency recorded on a separate frequency counter (Keysight 53220A), synchronised to the same TTL. A single sequence performs two laser frequency sweeps. The raw fluorescence trace is notch-filtered at \SI{50}{\hertz} to suppress mains noise. After which, the timing between the oscilloscope and frequency counter clocks is then aligned using the rising edge of the TTL pulse. Interpolation is used to map a beat-note frequency to every fluorescence sample. The beat note frequency $f_{\mathrm{beat~note}}$  is converted into a detuning, in units of natural linewidth $\Gamma$ by:
\begin{equation}
  \Delta_{\mathrm{beat~note}}  \;=\; \frac{2 f_{\mathrm{AOM}} - f_{\mathrm{beat~note}}}{\Gamma}
  \label{eq:detuning_scale}
\end{equation}
where $f_{\mathrm{AOM}}$ is the AOM frequency used for the polarisation-spectroscopy lock; the factor of two arises because the AOM is double-passed. The resulting spectra is then fitted using a Voigt profile, with the transverse velocity calculated using the Gaussian contribution ($\sigma_T$) from the fit parameter. Assuming the transverse velocity distribution follows a Maxwell-Boltzmann distribution, the transverse velocity is assumed to be the same as the most probable velocity and error, which can be expressed as~\cite{hughes2010measurements}:
\begin{equation} \label{eq:transverse_vel}
    v_{\perp,\mathrm{mp}} = \sqrt{\frac{2k_BT}{M}} = \frac{c}{\nu_0}\sigma_T
\end{equation}

where $k_B$ is the Boltzmann constant, T is the temperature of the atomic beam, M is the mass of Sr88, c is the speed of light and $\nu_0$ is the frequency for the $^1$S$_0$$-$$^1$P$_1$ transition.

The transverse velocities plotted in figure~\ref{fig:Atomic_beam_characteristic}(b) are the mean of three separate ($\langle v_{\perp,\mathrm{mp}}  \rangle$), with the reported error being the maximum spread of the data around the mean\cite{hughes2010measurements}:

\begin{equation}\label{eq:transverse_vel_err}
    \Delta v_{\perp,\mathrm{mp}} = \mathrm{max}\left( \mathrm{max}(v_{\perp,\mathrm{mp}}^{(\mathrm{i})}) - \langle v_{\perp,\mathrm{mp}}  \rangle , \langle v_{\perp,\mathrm{mp}}\rangle - \mathrm{min(v_{\perp,\mathrm{mp}}^{(\mathrm{i})})}\right)
\end{equation}

\subsection{Unidirectional probe scaling}\label{Appendix:single_probe_scaling}
A unidirectional transverse probe beam deviates from the simple two-level scattering-rate scaling used in the conversion factor. To correct the unidirectional probe ToF-derived flux distributions, a simulation-derived weighting factor is applied. Using Monte Carlo simulation outputs, the peak-detected fluorescence is extracted as a function of probe saturation for a unidirectional probe beam and fitted to a scattering-rate-like model. The ratio is then taken between the scattering rate equation ($\G_{\mathrm{sc}}(\langle s_{\textrm{probe}} \rangle)$) and the unidirectional probe beam fit ($m_{\mathrm{fit}}(\langle s_{\textrm{probe}} \rangle)$) as:
\begin{equation}
\eta(\langle s_{\textrm{probe}} \rangle)\equiv \frac{\G_{\mathrm{sc}}(\langle s_{\textrm{probe}} \rangle)}{m_{\mathrm{fit}}(\langle s_{\textrm{probe}} \rangle)},
\end{equation}
For the ToF measurements, the probe power is fixed at 12~mW. Using the average intensity for a unidirectional beam, the probe saturation parameter is $\langle s_{\textrm{probe}} \rangle = $~0.72. The resulting ratio is found to be,
\begin{equation}
\eta_{\mathrm{ToF}}(0.72) \approx 2.3
\end{equation}
This is the weight used to scale the unidirectional probe ToF-derived distributions shown in the main results,
\begin{equation}
\phi_{\mathrm{single}}^{\mathrm{(scaled)}}(v)=\eta_{\mathrm{ToF}}\;\phi_{\mathrm{single}}(v),
\Phi_{\mathrm{single}}^{\mathrm{(scaled)}}=\eta_{\mathrm{ToF}}\;\Phi_{\mathrm{single}}
\label{eq:single_probe_weighting}
\end{equation}

\subsection{Time-of-flight processing pipeline}\label{Appendix:tof_processing}
The fluorescence decay trace is synchronised to the rising edge of a TTL trigger for switching off the AOM coupled to the 2D MOT beams. The raw trace is then passed through a Butterworth low-pass filter (\SI{300}{\hertz} cut off), to suppress high-frequency noise which allows for a clear derivative signal $\big(-dV_{\textrm{PD}}/{d\textrm{t}}\big)$ to be obtained. The decay-time sample is converted to a longitudinal velocity of a known flight distance $\ell$, which is \SI{262.5}{\milli\meter} for this experiment.
\\~\\
The time-of-flight (ToF) analysis was performed on each repeat measurement individually before combining results at the dataset level. Each raw fluorescence trace was first aligned to the TTL trigger (rising) edge, then passed through a low-pass filter, and finally, the background noise was removed from the signal, resulting in the fluorescence decay in figure~\ref{fig:TOF_decay}. The time axis was then converted into longitudinal velocity using the known flight distance $\ell$,
\begin{equation}
  v_{\parallel} \;=\; \frac{\ell}{t},
  \qquad
  \Delta_{v_{\parallel}}
  \;=\; v_{\parallel}\sqrt{
        \left(\frac{\Delta_{\ell}}{\ell}\right)^{2}
        + \left(\frac{\Delta_{t}}{t}\right)^{2}},
  \label{eq:time_to_velocity}
\end{equation}

where the uncertainty on the propagation path ($\Delta\ell$) is the radius of the probe beam and the arrival time error ($\Delta t$) is the photodiode rise time.
\\~\\
The most probable longitudinal velocity ($v_{\parallel, \textrm{mp}}$) is then obtained by fitting to a Gaussian distribution,
\begin{equation}
f(v_{\parallel})=A\exp\!\left[-\frac{1}{2}\left(\frac{v_{\parallel}-v_{\parallel, \textrm{mp}}}{\sigma}\right)^2\right]
\label{eq:tof_gaussian}
\end{equation}
where $v_{\parallel, \textrm{mp}}$ is the most probable longitudinal velocity, $\sigma$ is the standard deviation, and $A$ is the amplitude. The most probable longitudinal velocity in figure~\ref{fig:Atomic_beam_characteristic}(b) is taken as the mean of three fitted most probable velocities ($\langle v_{\parallel,\mathrm{mp}}\rangle$), with the error being the maximum spread of about the mean~\cite{hughes2010measurements}:

\begin{equation}\label{eq:longitudinal_vel_err}
        \Delta v_{\parallel,\mathrm{mp}} = \mathrm{max}\left( \mathrm{max}(v_{\parallel,\mathrm{mp}}^{(\mathrm{i})}) - \langle v_{\parallel,\mathrm{mp}}  \rangle , \langle v_{\parallel,\mathrm{mp}}\rangle - \mathrm{min(v_{\parallel,\mathrm{mp}}^{(\mathrm{i})})}\right)
\end{equation}

Given the most probable transverse and longitudinal velocity, we can calculate the atomic beam divergence,
\begin{equation}
\theta_{\mathrm{div}}=\frac{v_{\perp,\mathrm{mp}}}{v_{\parallel,\mathrm{mp}}} 
\end{equation}
with uncertainty,
\begin{equation}
\Delta_{\theta}
=
\left|\theta_{\mathrm{div}}\right|
\sqrt{
\left(\frac{\Delta v_{\perp,\mathrm{mp}}}{v_{\perp,\mathrm{mp}}}\right)^2 
+
\left(\frac{\Delta v_{\parallel,\mathrm{mp}}}{v_{\parallel,\mathrm{mp}}}\right)^2}
\label{eq:divergence_err}
\end{equation}
\\~\\
Using the beam divergence, the cloud size at the fluorescence region can be estimated as,
\begin{equation}
d_{cloud}
=
\theta_{\mathrm{div}}\,\ell
\label{eq:cloud_size}
\end{equation}
with corresponding propagated uncertainty,
\begin{equation}
\Delta_{d}
=
\left|d_{cloud}\right|\sqrt{
\left(\frac{\Delta_{\ell}}{\ell}\right)^2
+
\left(\frac{\Delta_{\theta}}{\theta_{\mathrm{div}}}\right)^2 }
\end{equation}

The fitted Gaussian to the flux-per-velocity distribution $\phi(v_{\parallel})$ is integrated analytically to obtain the flux,
\begin{equation}
\Phi
=
\int_{v_{\parallel,\min}}^{v_{\parallel,\max}}
A\exp\!\left[-\frac{1}{2}\left(\frac{v_{\parallel}-v_{\parallel,\mathrm{mp}}}{\sigma}\right)^2\right]\,dv_{\parallel}
\end{equation}
which evaluates to
\begin{equation}
\Phi
=
A\sigma\sqrt{\frac{\pi}{2}}
\left[
\operatorname{erf}\!\left(\frac{v_{\parallel,\max}-v_{\parallel,\mathrm{mp}}}{\sqrt{2}\sigma}\right)
-
\operatorname{erf}\!\left(\frac{v_{\parallel,\min}-v_{\parallel,\mathrm{mp}}}{\sqrt{2}\sigma}\right)
\right]
\label{eq:gaussian_integral}
\end{equation}
where $v_{\parallel,\min}=0$ and $v_{\parallel,\max}=30~\mathrm{m\,s^{-1}}$ and $\sigma$ taken from the Gaussian best fit parameters. The flux is calculated individually for each ToF distribution repeat, with the flux values shown in figure~\ref{fig:Atomic_beam_characteristic}(a) being the average of three measurements ($\langle \Phi  \rangle$). The reported flux error is a combination of the statistical error calculated using the maximum spread about the mean ($\sigma_{\mathrm{stat}}$)~\cite{hughes2010measurements}, the systematic error ($\sigma_{\Phi,\mathrm{sys}}$) from the conversion factor and the total error ($\sigma_{\Phi,\mathrm{tot}}$) being the quadrature of the statistical and systematic error:
\begin{equation}
    \sigma_{\Phi,\mathrm{stat}}
    \;=\;
    \mathrm{max}\left(~\mathrm{max}\left(\Phi^{(\mathrm{i})}\right) - \langle \Phi  \rangle , \langle \Phi \rangle - \mathrm{min\left(\Phi^{(\mathrm{i})}\right)}~\right)
    \label{eq:flux_stat}
\end{equation}

\begin{equation}
    \sigma_{\Phi,\mathrm{sys}}
    \;=\;
    |\Phi|\;\frac{\Delta{C_{PD}}}{C_{\mathrm{PD}}}.
    \label{eq:flux_sys}
\end{equation}

\begin{equation}
    \sigma_{\Phi,\mathrm{tot}}
    \;=\;
    \sqrt{
      \sigma_{\Phi,\mathrm{stat}}^2
      +\sigma_{\Phi,\mathrm{sys}}^2
    }.
    \label{eq:total_err}
\end{equation}

\section{Spectra and ToF decay signal as a function of $s_{push}$}

\begin{figure*}[t]
    \includegraphics[width=\textwidth]{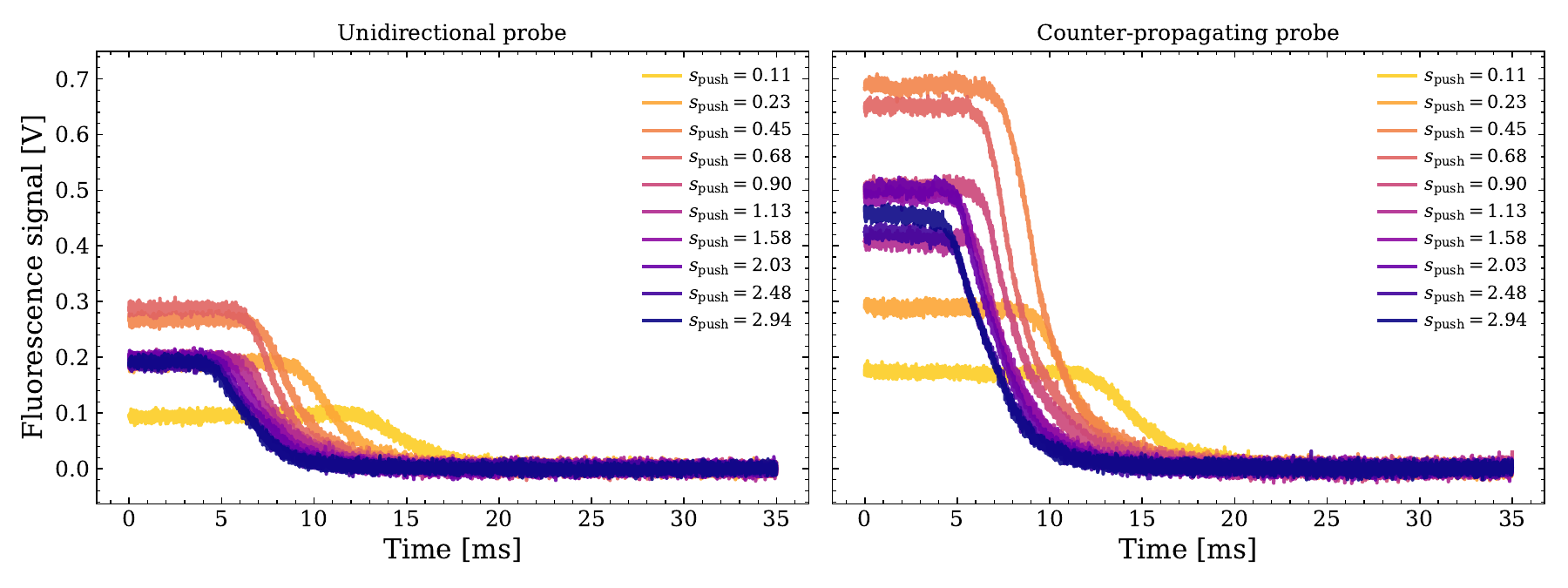}
 \caption{Experimental data monitoring the fluorescence decay observed in chamber 2 by switching the 2D~MOT beams off using an AOM, using a unidirectional probe beam (left) and counter-propagating probe beam (right) at $s_{\textrm{push}} = 0.11$ to $s_{\textrm{push}} = 2.94$}
 \label{fig:TOF_decay}
\end{figure*}

Table~\ref{beam_param} summarises beam parameters, and figures~\ref{fig:TOF_decay} shows the representative ToF traces across varying push saturation parameters.

\begin{table*}[!t]
\centering
\caption{\label{beam_param} Beam properties used in the experiment, with strontium atomic oven heated to \SI{400}{\celsius}.}
\resizebox{\textwidth}{!}{%
\begin{tabular}{|l|c|c|c|}
\hline
Beam Type & Beam power / mW & Beam diameter ($1/e^2$) / mm & Beam detuning / $\Gamma$ \\
\hline
2D MOT (per beam) & 35 & 30 & -1 \\
\hline
Push & \numrange{0.05}{1.3} & 1.7 & 0 \\
\hline
Probe (ToF) & 12 & 7.2 & 0 \\
\hline
Probe (spectroscopy) & \numrange{1}{10} & 7.2 & \numrange{-3}{3} \\
\hline
\end{tabular}%
}
\end{table*}

\FloatBarrier
\bibliography{references}

\end{document}